%% file: hc1.tex
\documentstyle[prb,aps]{revtex}
\begin{document}
\title{Local susceptibilities in semi-infinite antiferromagnet
chains: elementary perspectives}
\author{Martin P. Gelfand and Elisabeth F. Gl\"oggler}
\address{Department of Physics, Colorado State University, 
Fort Collins, Colorado 80523}
\maketitle

\begin{abstract}
Using conformal field theory methods Eggert and Affleck
\cite{eggert95} have shown that the semi-infinite $S=1/2$ Heisenberg 
antiferromagnetic chain exhibits a remarkable alternation in its 
local response to a uniform field at low temperatures.  
Such alternation is
not an essentially quantum effect: 
similar, and sometimes stronger,
susceptibility alternation is a feature of
classical Heisenberg-Ising chains at $T=0$.
In $S=1/2$ chains, susceptibility alternation
is not unique to the Heisenberg model, but can 
be seen in expansions about the Ising model 
and also in the $XY$ model.
\end{abstract}


\maketitle

\narrowtext

\section{Introduction}

In a system of coupled spins which lacks translational
invariance, the response to a {\it uniform\/} magnetic field 
also lacks translational invariance.  Labeling the
spins by a site index $n$ we can define zero-field
local susceptibilities
\begin{equation}
\chi^{\alpha\beta}_n = 
{\partial \langle S^\alpha_n\rangle \over \partial h^\beta}
\Big|_{{\bf h}={\bf 0}}
\end{equation}
(where $\alpha$ and $\beta$ denote Cartesian axes)
which will be the principal subject of this paper.

Eggert and Affleck\cite{eggert95} recently considered
the local susceptibilities in the semi-infinite $S=1/2$
Heisenberg antiferromagnet
\begin{equation}
{\cal H}= \sum_{n\ge1} J\,{\bf S}_n \cdot {\bf S}_{n+1}
                     + {\bf h}\cdot {\bf S}_n
\label{eq:halfchain}
\end{equation}
using powerful methods of boundary conformal field theory.
In this case spin-rotational symmetry implies
that $\chi_n^{\alpha\beta}= \delta_{\alpha,\beta} \chi_n$.
Their results have the following form, which we expect
to be asymptotically exact in the large-$n$  and 
low-temperature limit
(up to possible logarithmic corrections).  The local susceptibility
can be decomposed as $\chi_n=\chi^{\rm uni}_n + (-1)^{n-1} \chi^{\rm alt}_n$
where $\chi^{\rm uni}$ and $\chi^{\rm alt}$ are separately smooth,
positive functions of $n$.  The ``uniform'' part approaches
the susceptibility of an infinite chain with increasing $n$.
The ``alternating'' part is given by 
\begin{equation}
\chi^{\rm alt}_n \propto  n  
\left(\beta J^3 \sinh (4 n/\beta J)\right)^{-1/2} 
\label{eq:eggaff}
\end{equation}
with $\beta=k_B T$ and a proportionality constant of order unity.
This result has the remarkable feature that at $T=0$
the alternating part increases with $n$, and indeed it
diverges like $n^{1/2}$.

The local susceptibility in spin chains is a question of more
than purely theoretical interest.  It is the local susceptibility
which is relevant to local magnetic probes such as NMR, and
in any real sample of a quasi-one-dimensional magnet there
will be some defects that break the chains.  Indeed, the
effects of open chain ends may have already been seen in NMR studies
of Sr$_2$CuO$_3$.\cite{takigawa96}

The questions we would like to address are complementary
to those dealt with by Eggert and Affleck.  Is there
a {\it simple\/} way to understand the remarkable local
susceptibility alternation which they found?  
Is that alternation unique to Heisenberg models, or
is it a ubiquitous feature of semi-infinite antiferromagnetic
spin-chains?  We will explore these questions at $T=0$,
where the alternation phenomenon is strongest.

\section{Classical Spin-chains}
\label{sec:classical}

Eggert and Affleck suggested that the local susceptibility
alternation was a purely quantum mechanical effect.  Here
we will argue that, at least in the context of spin operators,
it is actually a classical effect.  Indeed, we will show
that classical version of the model (\ref{eq:halfchain})
has even stronger susceptibility alternation than the
$S=1/2$ quantum result (\ref{eq:eggaff}), and thus
quantum fluctuations, like thermal fluctuations,
act to suppress the susceptibility alternation.

Let us consider an $N$-site classical Heisenberg-Ising chain
\begin{equation}
{\cal H}= \sum_{n=1}^{N-1}
J_z S^z_n S^z_{n+1} + J_\perp(S^x_n S^x_{n+1} + S^y_n S^y_{n+1})
                     + {\bf h}\cdot \sum_{n=1}^{N}{\bf S}_n
\label{eq:HIham}
\end{equation}
where the spins are unit vectors.
We shall ultimately be interested in the $N\to\infty$ limit.  
It will prove convenient
to represent the spins in terms of their azimuthal angle $\phi$
and projection on the $xy$ plane $s\equiv \sin\theta$,
and to define
\begin{equation}
\lambda = J_\perp/J_z \ .
\end{equation}

If we are going to consider the $T=0$ local 
susceptibilities it should be evident that for Ising
anisotropy we should be examining the $xx$ (or $yy$)
susceptibilities while for $XY$ anisotropy the 
$zz$ susceptibilities are of interest.
In the limit that $\lambda=1$ 
the choice of susceptibility is irrelevant (so long as it
is diagonal) but for $\lambda<1$ 
the $zz$ susceptibility is very different in character.
For even $N$ there are two degenerate ground 
states in zero field (the "Ising states" in which
the spins are perfectly aligned with the $z$ axis and 
antiferromagnetically ordered) which remain the ground
states until a finite value of $h^z$ is achieved, so the
zero-field local susceptibilities are all strictly vanishing
whether one averages over both ground state configurations or not
in carrying out the $T=0$ average.
For odd $N$ the two Ising state energies vary linearly and cross
when $h^z=0$, so the natural characterization of the small-field
behavior of $\chi_n$ would be $(-1)^{n-1} 2 \delta(h)$ and
the zero-field limit would be ill-defined.  Clearly the
$N\to\infty$ limit is also ill-defined at $T=0$.
For $\lambda>1$ the $xx$ and $yy$ susceptibilities
are analogously pathological.
In contrast, the zero-temperature local transverse
susceptibilities $\chi^{xx}_n$ at $\lambda<1$
are well defined for all finite-length chains
and have smooth $N\to\infty$ limits;
and likewise for $\chi^{zz}_n$ at $\lambda>1$.  Let us
now calculate them for general $\lambda$ and $N$,
beginning with the case of Ising anisotropy.

For ${\bf h}=h\hat x$ the ground state has
alternating $S^z$ so that calculating the
ground state configuration involves minimizing 
\begin{eqnarray}
{\cal H} &=& \sum_{n=1}^{N-1} \Big(
-J_z\sqrt{(1-s_n^2)(1-s_{n+1}^2)} \nonumber\\
&+& J_\perp s_n s_{n+1} \cos(\phi_n-\phi_{n+1}) \Big)
- h\sum_{n=1}^N s_n \cos\phi_n \ .
\label{eq:eclasis}
\end{eqnarray}
Setting 
$\partial {\cal H}/\partial \phi_n= \partial {\cal H}/\partial s_n = 0$ 
leads to the conditions
\begin{eqnarray}
&-&h s_n \sin\phi_n /J_\perp \nonumber\\ 
&=&  s_{n-1}s_n \sin(\phi_{n-1}-\phi_n)
             +s_n s_{n+1} \sin(\phi_n-\phi_{n+1}) 
\label{eq:phiextreme}
\end{eqnarray}
and
\begin{eqnarray}
h&& \cos\phi_n/J_z
 = s_n {\sqrt{1-s_{n-1}^2} + \sqrt{1-s_{n+1}^2}\over\sqrt{1-s_{n}^2}}
\nonumber\\
 +&& \lambda(s_{n-1}\cos(\phi_{n-1}-\phi_n) + s_{n+1}\cos(\phi_n-\phi_{n+1}))
\label{eq:sextreme}
\end{eqnarray}
for $n=2\ldots N-1$. For $n=1$ ($N$)
the terms containing $s_0$ and $\phi_0$ ($s_{N+1}$ and $\phi_{N+1}$)
are dropped but the equations are otherwise identical.
The equations (\ref{eq:phiextreme}) have the solution
$\phi_n=0$ for all $n$.  This makes perfect sense:  why should
the spins deviate from the $xz$ plane?  
(Note that $\phi_n=0$ for all $n$ will only correspond to the
minimum energy configuration when $h>0$; for $h<0$, $\phi_n=\pi$
gives the minimum.  There are also solutions to (\ref{eq:phiextreme})
in which some of the $\phi_n$ are 0 and the others are $\pi$,
but these are not energy minima.)
Then in (\ref{eq:sextreme})
all of the cosines can be replaced by 1, and $s=S^x$.
To obtain the zero-field local susceptibilities 
we linearize (\ref{eq:sextreme}) with respect to the $s_n$ and obtain
(with $\chi\equiv\chi^{zz}$)
\begin{equation}
2\chi_n + \lambda (\chi_{n-1} + \chi_{n+1}) = 1/J_z
\label{eq:chiformula}
\end{equation}
for $n=2\ldots N-1$, and the boundary conditions
\begin{equation}
\chi_1 + \lambda \chi_2  = 1/J_z \qquad
\chi_N+ \lambda \chi_{N-1}= 1/J_z \ .
\label{eq:chibc}
\end{equation}

Before we proceed to discuss the solution of the linear
system which governs the $\chi_n$ let us consider what happens
for $XY$ anisotropy and fields ${\bf h}=h \hat z$.
In this case the ground state will have non-alternating $S^z$
so instead of (\ref{eq:eclasis}) the quantity to minimize
will have a plus rather than minus sign associated with
the term $\sqrt{(1-s_n^2)(1-s_{n+1}^2)}$.  In addition, the
external field term $h s_n \cos\phi_n$ is replaced by 
$h \sqrt{1-s_n^2}$.  Then the energy-minimizing solution
to the analog of Eq.~(\ref{eq:phiextreme}) is $\phi_n-\phi_{n+1}=\pi$.
There is clearly degeneracy in the ground state:  starting
from any spin configuration, all
spins can be rotated by an equal amount about the $z$
axis without changing the energy.  Then, defining $c=\cos\theta=S^z$,
the analog of Eq.~(\ref{eq:sextreme}) becomes (extremalizing the
energy with respect to the $c_n$ rather than the $s_n$)
\begin{equation}
h/J_z=c_{n-1}+c_{n+1} -  \lambda c_n {\sqrt{1-c_{n-1}^2}+\sqrt{1-c_{n+1}^2}
\over \sqrt{1-c_n^2} }
\end{equation}
which upon linearization with respect to the $c_n$ leads to
the relation 
\begin{equation}
2\lambda\chi_n +  \chi_{n-1} + \chi_{n+1} = 1/J_z
\end{equation}
where $\chi=\chi^{zz}$ here.  This has the same form as
Eq.~(\ref{eq:chiformula}), if in the former one makes the
substitutions $\lambda \to 1/\lambda$ and $J\to \lambda J$.
In other words, the equations are related by the interchange
$J_z \leftrightarrow J_\perp$.  So,
solving Eq.~(\ref{eq:chiformula}) for the transverse
local susceptibilities in Ising-anisotropic systems is enough:
there is no new calculation to be done for $XY$-anisotropic
systems.

Now let us get on with the business of solving 
for the local susceptibilities.
Two special cases are of particular interest.  

For the Ising model, $\lambda=0$,
one has $\chi_n=1/(2J_z)$ for all interior sites on the chain,
while $\chi_1=\chi_N=1/J_z$.  Note there is no alternation
in this case, and for the interior sites the response is
the same as if there were no boundaries at all.

For the Heisenberg model, 
$\lambda=1$ and  $J_z=J_\perp\equiv J$,
the behavior of the system
depends markedly on the parity of $N$.   If $N$ is odd, then
the ground-state energy varies linearly in $|h|$ because
the ground-state configuration consists of aligning
the odd-numbered sites with the field and anti-aligning
the even-numbered sites.  Therefore the local susceptibility
has $\delta$-function character at every site, with amplitude proportional
to $(-1)^{n-1}$.  This corresponds to the greatest
conceivable local susceptibility alternation. 
For even $N$, the spins instead cant in response to a field,
and the local susceptibilities are well defined.
However, the linear system (\ref{eq:chiformula}) and
(\ref{eq:chibc}) does not have a unique solution:
the linear operator has a zero-mode of the form
$(1,-1,1,-1,\ldots,1,-1)$.  This apparent degeneracy 
of spin configurations in the small-$h$ limit is
resolved by the fourth-order terms in the $s_i$
in the Hamiltonian.  One finds then that the $\chi_n$
satisfy $\chi_{N+1-n}=\chi_n$ and 
are conveniently described by the recurrence
relation $J\chi_{n+1}=J\chi_{n}+(-1)^n({1\over2}N-n)$
with the starting value $J\chi_1=N/4$.  For example,
for $N=8$ the sequence of $J\chi_n$ is (2,-1,1,0,0,1,-1,2).
There is certainly susceptibility alternation here, but
unlike the semi-infinite quantum chain the alternation
is largest near the boundaries rather than in the interior.
Despite the very different characteristics of the local
susceptibility for odd and even $N$, it seems that the
$N\to\infty$ limit for $\chi_n$ at fixed $n$ is well
described by $(-)^{n-1}\infty$: a striking result!

Let us now turn to the general problem of calculating
$\chi^{xx}_n$ for $0<\lambda<1$ and general $N$, with
particular attention to the $N\to\infty$ limit.
We have employed a generating-function technique,
which involves extending the $n$ values under consideration
from zero to infinity, defining $X(z)=\sum_{n\ge0}^\infty \chi_n z^{-n}$.
This sort of generating function is conventionally
known as a ``$z$-transform''.\cite{korn}
In order to satisfy (\ref{eq:chiformula}) when $n=1$
as well as the lower boundary condition in (\ref{eq:chibc})
the relation $\chi_0=-\chi_1/\lambda$ must hold; the
value of $\chi_0$ (and thence all the $\chi_n$) will
be uniquely determined later, in order to satisfy
the upper boundary condition.  Then we can apply
the standard result for $z$-transform solution
of linear recurrences:
\begin{equation}
X(z)={R(z)+I(z) \over Q(z)}
\end{equation}
where $Q(z)$ encodes the homogeneous terms in the recurrence
(and equals $z^2+2 \lambda^{-1} z +1$ in the present case),
$R(z)$ encodes the inhomogeneous terms (and equals
$[J_z \lambda (1-z)]^{-1}$ here) and $I(z)$ encodes
the initial conditions (and equals 
$\chi_0 z^2 + (\chi_1 + 2\lambda^{-1}\chi_0)z
= \chi_0 Q -\chi_0+\chi_1 z$ here).  In order to carry out the back-transform
for $X$ it is useful to introduce a partial-fraction decomposition
of $1/Q$, which makes it useful to denote the roots of
$Q$ by $z_\pm=\lambda^{-1}(-1\pm\sqrt{1-\lambda^2})$.
Then straightforward algebra, combined with the knowledge
that the $z$-transform of $(a^n-b^n)/(a-b)$ is $z/(z-a)(z-b)$
and also the relation between $\chi_0$ and $\chi_1$ given above,
leads to 
\begin{eqnarray}
\chi_n=&&{1\over \lambda J_z(z_+ - z_-)}
\left[ {1-z_+^n \over 1-z_+} - {1-z_-^n \over 1-z_-} \right]
\nonumber\\
&&-{1\over2}\lambda\chi_0(z_+^{n-1}+z_-^{n-1}) \ .
\label{eq:chimostsoln}
\end{eqnarray}
Inserting this expression into the upper boundary condition
yields the explicit expression for $\chi_0$ 
\widetext
\begin{equation}
\chi_0=
{-z_+^{N-2}(\lambda+(1-\lambda)z_+-z_+^2)
+z_-^{N-2}(\lambda+(1-\lambda)z_--z_-^2) 
+(1+\lambda)(z_+ - z_-)
\over 2J_z(1+\lambda)^2(1-\lambda)(z_+^{N-1}-z_-^{N-1})} \ .
\end{equation}
\narrowtext
In taking the $N\to\infty$ limit we apply the fact that
$|z_+|<1$ and $|z_-|>1$, giving 
\begin{eqnarray}
\lim_{N\to\infty}\chi_0 = &&
{1\over \lambda^2(1+\lambda)}
\Bigg[{\sqrt{1-\lambda^2}\over 1+\lambda}
\left({1\over\lambda}+{1\over2}\right) \nonumber\\
&&-\left({1\over\lambda}-{1\over2}\right)
-{1\over\sqrt{1-\lambda^2}} \Bigg] \ .
\end{eqnarray}
In conjunction with Eq.~(\ref{eq:chimostsoln}), this gives
an explicit solution for $\chi_n$ for the semi-infinite
chain.

The solution at hand is explicit but rather unwieldy.
It turns out that for $\lambda$ not too close to $1$
and $n$ not too large it is more convenient to apply 
numerical methods directly to the linear system
for the $\chi_n$.  Numerical solution is especially
easy in this case because the problem is a tridiagonal
linear system.  In Fig.~\ref{fig:clasplt} we present
results for $\chi_1$ through $\chi_6$; although the
calculations were done for $N=100$,
on the scale of this plot any finite-$N$ effects
are invisible.  Also shown, as the solid line, is
the susceptibility of a uniform chain.
Some key features of the results for $\lambda<1$
should be evident: the amount of alternation
decreases as one moves towards the chain interior,
with the rate of decrease depending on $\lambda$
(smaller $\lambda$ giving a stronger decrease), and
as $\lambda\to1$ all of the $\chi_n$ diverge.
What cannot be easily seen in the numerical solution
are the details of the behavior near $\lambda=1$,
since the finite $N$ effects are strongest there.
For this reason it is informative
to construct an expansion for $\chi_n$ in powers
of $\delta\equiv 1-\lambda$, using the explicit solution.
The leading terms are found to be
\begin{equation}
\chi_n \approx (-)^{n-1}2^{-3/2}\delta^{-1/2}
 + (1 + 2(n-1)(-)^n)/4 
\end{equation}
with the next term $O(\delta^{1/2})$.
Note that up to the order displayed $\chi_2=-\chi_3$,
$\chi_4=-\chi_5$, and so forth.  This behavior is quite
evident in the numerical solutions for $\delta\approx10^{-3}$.

Parenthetically, we should note that there is some similarity
between the problem that has been solved in this section,
and a class of problems that has arisen in the context
of magnetic multilayer systems, namely the determination
of energy (or free energy) minimizing spin configurations
which have variation only in one dimension.%
\cite{nortemann92,camley93}  Our problem is actually
much simpler. By restricting our attention to
the linear susceptibility the calculation reduces
to solution of a linear system, which allows for
a straightforward, if messy, analytic solution.
It is also the case that for magnetic multilayers
the interest lies in systems with large ``single-ion'' 
(where an ion corresponds to a magnetic layer) $XY$ anisotropy
and longitudinal magnetic fields, so there has been
no overlap with the results of the present work.

\section{Quantum spins chains}

Let us now return to $S=1/2$ quantum spin chains, and
consider whether the alternation in the $T=0$ local susceptibility
of the semi-infinite Heisenberg chains persists
when the rotational symmetry of the interaction in broken.
That is, we will consider again the Hamiltonian (\ref{eq:HIham})
but reinterpret the ${\bf S}_n$ as quantum spins rather
than classical spins.
For the $XY$ model ($J_z=0$), 
it is possible to exactly calculate
$\chi_n^{zz}$.  An exact calculation of $\chi_n^{xx}$ is
trivially possible for the Ising model, 
and furthermore one can construct 
perturbation expansions for the $\chi_n^{xx}$ in powers
of $\lambda$.  We will find a susceptibility alternation
for the $XY$ semi-infinite chain, albeit not so strong as 
for the Heisenberg model. For $\lambda < 1$ we will find behavior 
reminiscent of the classical Heisenberg-Ising semi-infinite chain,
except that as $\lambda\to1$ the local susceptibilities diverge
classically but we know they are finite quantum-mechanically.

\subsection{$XY$ Model}

For open chains it is straightforward to apply the
Jordan-Wigner transformation \cite{mattis} to transform the $XY$ model
with a field ${\bf h}=h \hat z$ into a model of noninteracting
spinless fermions, for 
which the corresponding Hamiltonian has the form
\begin{equation}
{\cal H}=J_\perp \sum_{i=1}^{N-1} (c^\dagger_i c_{i+1} + c^\dagger_{i+1} c_i)
- h \sum_{i=1}^N (c^\dagger_i c_i - {\textstyle 1\over2}) \ .
\end{equation}
The ground state is constructed by filling
all of the negative energy single-particle states.
The single-particle eigenvalues have the form $\lambda_m=-h+2J_\perp\cos q_m$
where $q_m=\pi m/(N+1)$ and $m$ runs over the integers $1 \ldots N$;
and the normalized eigenfunctions are given by 
$\psi_m(n)=\sqrt{2/(N+1)}\sin q_m n$.

The local magnetization is given by
\begin{equation}
\langle S^z_n \rangle =-{\textstyle 1\over2}+
\sum_{\{m|\lambda_m < 0\}} |\psi_m(n)|^2 \ ,
\label{eq:SzXYsum}
\end{equation}
but is not uniquely defined if there is an eigenvalue (corresponding
to $m=m_0$) which is exactly zero since in that case an infinitesimal
change in $h$ can cause a finite change in $\langle S^z_n \rangle$,
of $|\psi_{m_0}(n)|^2$.  

At $h=0$, for finite-length chains the parity of $N$ strongly
affects the susceptibilities.  For even $N$, the single-particle
eigenvalues have a gap of $O(N^{-1})$ around zero, and so
the zero-field local susceptibilities equal zero.
The zero-field local magnetizations
also vanish.
For odd $N$, $\lambda_{(N+1)/2}=0$; since the corresponding
wave vector is $\pi/2$ the susceptibility has $\delta$-function
character (with strength of $O(N^{-1})$) at $h=0$ on odd-numbered
sites and vanishes on even-numbered sites.  
(These peculiarities of of chains with open boundary conditions
were also noted, in a slightly different context, by 
S{\o}rensen and Affleck.\cite{sorensen96})

However, it turns out
that thermodynamic limit is perfectly well behaved 
despite the disparate characteristics of different parity
finite-length chains.  In the $N\to\infty$ limit one can
replace the sum in (\ref{eq:SzXYsum}) by an integral, as follows:
\begin{eqnarray}
\langle S^z_n(h)\rangle&& = -{1\over2}+{2\over N} 
\sum_{m=(N+1)({1\over2} - {h\over 2\pi J_\perp})}^N \sin^2 q_m 
\nonumber\\
&&\rightarrow
{2\over\pi}\int_{\pi/2-h/2J_\perp}^{\pi/2} dq\,\sin^2 q \ .
\end{eqnarray}
(In the sum there should be a
greatest-integer function acting on the $h$-dependent term
in the lower limit, however, that has no effect when
$N\to\infty$.)
Differentiation with respect to $h$ leads to
\begin{equation}
\chi^{zz}_n=(\pi J_\perp)^{-1}\sin^2 {\pi n\over2} \ ,
\end{equation}
that is, $\chi^{zz}_n$ vanishes on even-numbered sites and takes the
value $(\pi J_\perp)^{-1}$ on odd-numbered sites.
This certainly constitutes an alternation in the local
susceptibility, one which is weaker than in the quantum
Heisenberg model but, surprisingly, stronger than in
the classical $XY$ model.

\subsection{Expansions about the Ising Model}

At present exact results for the $T=0$ local transverse susceptibilities
of semi-infinite $S=1/2$ chains are only available for
the Ising and $XY$ models.  In order to obtain unbiased approximations
one might consider carrying out 
exact diagonalization or quantum Monte Carlo calculations 
on open $N$-site chains.  
Here we will take another approach which avoids
the need for an explicit $N\to\infty$ limit, namely, 
the calculation and extrapolation of high-order convergent
perturbation expansions about the Ising model.
Such expansions have proven very informative for
infinite, homogeneous systems in both one
and two dimensions;\cite{isingexps} for more general
discussions of the techniques used in high-order convergent
perturbation expansions see Ref.~9.
We present here the first results obtained using such methods
for an inhomogeneous system.  The calculations were carried
out by R. R. P. Singh.\cite{singhprivcomm}

Expansions were calculated for 
$J_z\chi_n$, with $n=1\ldots 15$, up to
order $\lambda^{14}$ (recall $\lambda=J_\perp/J_z$).
Henceforth let us set $J_z\equiv1$ for convenience.
The series for $\chi_n$ is identical to the series
for the bulk $\chi$ up to order $\lambda^{n-2}$,
and so it is more informative to consider
the series for $\Delta\chi_n\equiv \chi_n - \chi$,
for which the first nonzero term is at order $\lambda^{n-1}$.
A subset of the calculation series coefficients,
sufficient for our present purposes, is presented
in Table~\ref{table:chinser}.

The most prominent features of the $\Delta\chi_n$ series
that should be evident in the Table are (i)
the aforementioned vanishing of terms up to
order $\lambda^{n-2}$, and (ii) the sign of the
leading term alternates between positive, for odd $n$,
and negative, for even $n$.
In fact, the same properties hold for
the semi-infinite {\it classical\/} Heisenberg-Ising model,
as one can verify by carrying out the Taylor expansion
in $\lambda$ of the exact classical solution presented
in Sec.~\ref{sec:classical}. 
The leading term in $\Delta\chi_n$ for the classical
model is $(-\lambda/2)^{n-1}/2$.
Note that the coefficients in that case are
{\it smaller\/} in magnitude than the corresponding
coefficients for the $S=1/2$ system and the
difference increases with $n$, so the
susceptibility alternation near the Ising limit
is apparently {\it enhanced\/} by quantum fluctuations.

Given the reasonably long series available
one might hope to
obtain reliable estimates of the $\chi_n$ over the
whole interval $[0,1]$.
However, series extrapolation turns out to be 
rather difficult.  
Consider first the bulk susceptibility.  Direct
Pad\'e approximants appear to be well-behaved,
and suggest that $\chi$ for the Heisenberg model
is in the range $0.107$ to $0.111$.  In fact the
exact value is $0.1013\ldots$ \cite{griffiths}.
The Pad\'e approximants are systematically in error,
albeit modestly, presumably because  there
is a strong singularity in $\chi$ as $\lambda\to(-1)^+$  and 
a weak singularity as $\lambda\to1^{-}$ \cite{chising}.
It is possible that related singularities
might exist in all the $\Delta\chi_n$, although
we have not considered that problem in detail.  
A more substantial barrier to reliable extrapolations 
based on the $\Delta\chi_n$ series is that, in comparison 
with the bulk $\chi$ series, they are quite irregular and
the degree of irregularity increases with $n$.
Only for $\Delta\chi_1$ do we dare offer an estimate
of its value at $\lambda=1$:  a broad interval from
0.7 to 1.3.  In fact the approximants are clustered
at the ends of the interval rather than being spread
uniformly, so an estimate presented in the form $A\pm B$
would be somewhat misleading.

\section{Conclusions}

The phenomenon of local transverse-susceptibility alternation
in semi-infinite spin-chains at $T=0$ turns out to be quite
rich.  For {\it classical\/} spins, the difference
between the local susceptibility and the susceptibility
of a uniform chain decreases with distance from the chain
end at fixed anisotropy, but diverges as the anisotropy
approaches zero at any fixed distance.  For classical
Heisenberg chains at $T=0$ the local susceptibilities
are all divergent and alternating in sign; thus quantum
fluctuations in isotropic $S=1/2$ chains have the effect 
of {\it suppressing\/} susceptibility alternation, most 
effectively near the end of the chain.  
These findings lead us to suggest that 
the susceptibility alternation in semi-infinite
$S=1/2$ Heisenberg chains \cite{eggert95} is
not an essentially quantum phenomenon.
They might also lead one to conclude
that quantum fluctuations always suppress the local susceptibility
alternation in semi-infinite chains, but that turns
out to be incorrect.
For both the $XY$ model and near the Ising limit,
the $S=1/2$ chain has stronger alternation than
the classical chain.

We thank I. Affleck and R. R. P. Singh for discussions,
and are especially grateful to Prof.~Singh for allowing
us to present his unpublished series expansions.
This work was supported by the National Science Foundation
though Grant DMR 94--57928 (MPG) and by the
Gottlieb Daimler~-~und Karl Benz~-~Stiftung through
the ``Forschungsarbeit im Ausland'' program (EFG).

\begin{figure}
\caption{Plot of the local susceptibilities $\chi_n$
for $n=1$ through 6 for the classical Heisenberg-Ising
semi-infinite chain, as a function of the parameter
$\lambda=J_\perp/J_\parallel$.  The solid line
is the bulk susceptibility.}
\label{fig:clasplt}
\end{figure}

\mediumtext
\input table

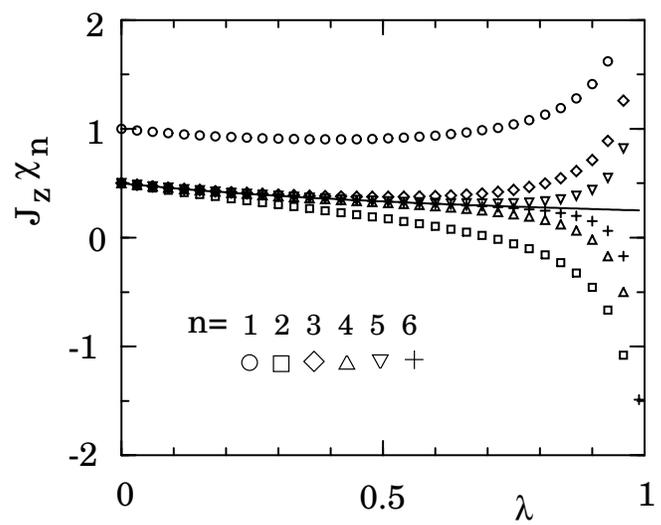
\end{document}

%% file: table.tex
\begin{table}
\caption{}
\begin{tabular}{cddddd}
 $n$ & $\chi$ & $\Delta\chi_1$ & $\Delta\chi_2$ & 
                $\Delta\chi_3$ & $\Delta\chi_4$ \\
\hline
 0 &    0.5        &    0.5       &    0.        &
                        0.        &    0.        \\

 1 & $-$1.         & $-$0.5       & $-$1.        &
                        0.        &    0.        \\

 2 &    1.5        &    1.        & $-$0.5       &
                        3.125     &    0.        \\

 3 & $-$2.25       & $-$0.75      &    1.9375    &
                     $-$2.5       & $-$6.3125    \\

 4 &    3.28125    &    0.53125   & $-$0.6354167 &
                     $-$1.1197917 &    0.8125    \\

 5 & $-$4.53125    & $-$0.28125   &    0.3645833 &
                     $-$0.4010417 &   16.6875    \\
 
 6 &    6.0976563  &    0.1210937 & $-$3.0285012 &
                        2.4624928 & $-$3.9049479 \\
 
 7 & $-$8.0390625  &    1.140625  &    4.5787037 &
                     $-$0.5331670 & $-$18.5109592 \\
 
 8 &   10.3918457  & $-$2.7512207 & $-$4.5152543 &
                        0.8260136 & $-$0.1684826 \\

\end{tabular}
\label{table:chinser}
\end{table}

%% file: hc1.bbl
\begin{references}

\bibitem{eggert95}
S. Eggert and I. Affleck,
Phys. Rev. Lett. {\bf 75}, 934 (1996).

\bibitem{takigawa96}
M. Takigawa, N. Motoyama, H. Eisaki, and S. Uchida,
Phys. Rev. Lett. {\bf 76}, 4612 (1996).

\bibitem{korn}
G. A. Korn and T. M. Korn,
{\sl Mathematical Handbook for Scientists and Engineers},
(McGraw-Hill, New York, 1961), pp. 648--650.

\bibitem{nortemann92}
F. C. N{\" o}rtemann, R. L. Stamps, A. S. Carri{\c c}o, and R. E. Camley,
Phys. Rev. B. {\bf 46}, 10847 (1992).

\bibitem{camley93}
R. E. Camley and R. L. Stamps,
J. Phys. Cond. Matt. {\bf 5}, 3727 (1993).

\bibitem{mattis}
See, for example,
D. C. Mattis, {\sl The Theory of Magnetism I},
Berlin: Springer-Verlag, 1981.  

\bibitem{sorensen96}
E. S. S{\o}rensen and I. Affleck,
Phys. Rev. B {\bf 53}, 9153 (1996);
see in particular Sec.~V of this paper.

\bibitem{isingexps}
See, for example, 
R. R. P. Singh,
Phys. Rev. B {\bf 39}, 9760 (1989);
C. J. Hamer, W. H. Zheng, and J. Oitmaa,
Phys. Rev. B {\bf 50}, 6877 (1994).

\bibitem{exprevs}
M. P. Gelfand, R. R. P. Singh, and D. A. Huse,
J. Stat. Phys. {\bf 59}, 1093 (1990);
H.-X. He, C. J. Hamer, and J. Oitmaa,
J. Phys. A. {\bf 23}, 1775 (1990).

\bibitem{singhprivcomm}
R. R. P. Singh (private communication).

\bibitem{griffiths}
R. B. Griffiths,
Phys. Rev {\bf 133}, A768 (1964).

\bibitem{chising}
Since $\lambda=-1$ corresponds to a {\it ferromagnetic\/}
Heisenberg chain, $\chi$ will
diverge on approach to that point.  
That is responsible for the sign alternation
and the increase in the
magnitude of the series coefficients with order,
and complicates extrapolation of the series to
$\lambda=1$.
For $\lambda\to1^-$
we expect there to be a weak singularity
in the susceptibility since the antiferromagnetic
Heisenberg chain is critical. Its precise character
is immaterial, since unless the associated critical
exponent is unity it cannot be fully accounted for by 
Pad\'e approximants, but we believe it is a square-root
cusp.  (That is, we suppose the transverse susceptibility
to exhibit the same singularity as the spin stiffness at 
the Kosterlitz-Thouless transition.)
In fact, a plot of Euler sums at $\lambda=1$ up 
to $n$th order against $n^{-1/2}$ is nearly linear for $n>9$; and
extrapolating to $n=\infty$ leads to an estimate for
$\chi$ in the Heisenberg model of $0.1007(9)$, consistent
with the exact result.

\end{references}
